\documentstyle[12pt,epsf]{article}

\catcode`\@=11
\long\def\@makefntext#1{
\protect\noindent \hbox to 3.2pt {\hskip-.9pt  
$^{{\ninerm\@thefnmark}}$\hfil}#1\hfill}		

\def\@makefnmark{\hbox to 0pt{$^{\@thefnmark}$\hss}}  
	
\def\ps@myheadings{\let\@mkboth\@gobbletwo
\def\@oddhead{\hbox{}
\rightmark\hfil\ninerm\thepage}   
\def\@oddfoot{}\def\@evenhead{\ninerm\thepage\hfil
\leftmark\hbox{}}\def\@evenfoot{}
\def\sectionmark##1{}\def\subsectionmark##1{}}

\renewcommand{\thefootnote}{\fnsymbol{footnote}}

\newcounter{sectionc}\newcounter{subsectionc}\newcounter{subsubsectionc}
\renewcommand{\section}[1] {\vspace*{0.6cm}\addtocounter{sectionc}{1} 
\setcounter{subsectionc}{0}\setcounter{subsubsectionc}{0}\noindent 
	{\normalsize\bf\thesectionc. #1}\par\vspace*{0.4cm}}
\renewcommand{\subsection}[1] {\vspace*{0.6cm}\addtocounter{subsectionc}{1} 
	\setcounter{subsubsectionc}{0}\noindent 
	{\normalsize\it\thesectionc.\thesubsectionc. #1}\par\vspace*{0.4cm}}
\renewcommand{\subsubsection}[1]
{\vspace*{0.6cm}\addtocounter{subsubsectionc}{1}
	\noindent {\normalsize\rm\thesectionc.\thesubsectionc.\thesubsubsectionc. 
	#1}\par\vspace*{0.4cm}}

\newcounter{appendixc}
\newcounter{subappendixc}[appendixc]
\newcounter{subsubappendixc}[subappendixc]

\renewcommand{\appendix}[1] {\vspace*{0.6cm}
        \refstepcounter{appendixc}
        \setcounter{figure}{0}
        \setcounter{table}{0}
        \setcounter{equation}{0}
        \renewcommand{\thefigure}{\Alph{appendixc}.\arabic{figure}}
        \renewcommand{\thetable}{\Alph{appendixc}.\arabic{table}}
        \renewcommand{\theappendixc}{\Alph{appendixc}}
        \renewcommand{\theequation}{\Alph{appendixc}.\arabic{equation}}
        \noindent{\bf Appendix \theappendixc #1}\par\vspace*{0.4cm}}

\def\abstracts#1{{
	\centering{\begin{minipage}{12.2truecm}\footnotesize\baselineskip=12pt\noindent
	\centerline{\footnotesize ABSTRACT}\vspace*{0.3cm}
	\parindent=0pt #1
	\end{minipage}}\par}} 

\renewenvironment{thebibliography}[1]
	{\begin{list}{\arabic{enumi}.}
	{\usecounter{enumi}\setlength{\parsep}{0pt}
\setlength{\leftmargin 1.25cm}{\rightmargin 0pt}
	 \setlength{\itemsep}{0pt} \settowidth
	{\labelwidth}{#1.}\sloppy}}{\end{list}}

\newcounter{itemlistc}
\newcounter{romanlistc}
\newcounter{alphlistc}
\newcounter{arabiclistc}

\newcommand{\fcaption}[1]{
        \refstepcounter{figure}
        \setbox\@tempboxa = \hbox{\footnotesize Fig.~\thefigure. #1}
        \ifdim \wd\@tempboxa > 6in
           {\begin{center}
        \parbox{6in}{\footnotesize\baselineskip=12pt Fig.~\thefigure. #1}
            \end{center}}
        \else
             {\begin{center}
             {\footnotesize Fig.~\thefigure. #1}
              \end{center}}
        \fi}

\newcommand{\tcaption}[1]{
        \refstepcounter{table}
        \setbox\@tempboxa = \hbox{\footnotesize Table~\thetable. #1}
        \ifdim \wd\@tempboxa > 6in
           {\begin{center}
        \parbox{6in}{\footnotesize\baselineskip=12pt Table~\thetable. #1}
            \end{center}}
        \else
             {\begin{center}
             {\footnotesize Table~\thetable. #1}
              \end{center}}
        \fi}

\def\@citex[#1]#2{\if@filesw\immediate\write\@auxout
	{\string\citation{#2}}\fi
\def\@citea{}\@cite{\@for\@citeb:=#2\do
	{\@citea\def\@citea{,}\@ifundefined
	{b@\@citeb}{{\bf ?}\@warning
	{Citation `\@citeb' on page \thepage \space undefined}}
	{\csname b@\@citeb\endcsname}}}{#1}}

\newif\if@cghi
\def\cite{\@cghitrue\@ifnextchar [{\@tempswatrue
	\@citex}{\@tempswafalse\@citex[]}}
\def\citelow{\@cghifalse\@ifnextchar [{\@tempswatrue
	\@citex}{\@tempswafalse\@citex[]}}
\def\@cite#1#2{{$\null^{#1}$\if@tempswa\typeout
	{IJCGA warning: optional citation argument 
	ignored: `#2'} \fi}}

 1
 1
 1

\font\ninerm=cmr9

\begin{document}
\rightline{\vbox{\halign{&#\hfil\cr
&ANL-HEP-CP-95-85\cr
&November 22, 1995\cr}}}
\vspace{1in}
\centerline{\normalsize\bf TOP QUARK PRODUCTION DYNAMICS IN QCD
\footnote{Invited paper presented by E. L. Berger at the International Symposium
on Heavy Flavor and Electroweak Theory, Beijing, August 16 - 19, 1995}}
\baselineskip=22pt

\centerline{\footnotesize EDMOND L. BERGER}
\baselineskip=13pt
\centerline{\footnotesize\it High Energy Physics Division, Argonne National
Laboratory, Argonne, IL 60439-4815, USA}
\centerline{\footnotesize E-mail: ELB@hep.anl.gov}
\vspace*{0.3cm}
\centerline{\footnotesize and}
\vspace*{0.3cm}
\centerline{\footnotesize HARRY CONTOPANAGOS}
\baselineskip=13pt
\centerline{\footnotesize\it High Energy Physics Division, Argonne National
Laboratory, Argonne, IL 60439-4815, USA}
\centerline{\footnotesize E-mail: CONTOPAN@hep.anl.gov}
\vspace*{0.9cm}
\abstracts{A calculation of the total cross section for top quark production in
hadron-hadron collisions is presented based on an all-orders perturbative
resummation of initial-state gluon radiative contributions to the basic quantum
chromodynamics subprocesses.  Principal-value resummation is used to evaluate
all relevant large threshold contributions.  In this method there are no
arbitrary infrared cutoffs, and the perturbative regime of applicability is
well defined, two attributes that significantly reduce the estimated 
uncertainty of the results. For $p\bar{p}$ collisions at center-of-mass energy
$\sqrt{s}=1.8$ TeV and a top mass of 175 GeV, we obtain
$\sigma(t\bar{t})=5.52^{+0.07}_{-0.45} pb$, in agreement with experiment. 
Predicted cross sections are provided as
a function of top mass in $p\bar{p}$ collisions at $\sqrt{s}=2.0$ TeV and
in $p p$ collisions at CERN LHC energies.}

\normalsize\baselineskip=15pt
\setcounter{footnote}{0}
\renewcommand{\thefootnote}{\alph{footnote}}
\section{Introduction}

The quest for the top quark $t$ reached fruition in the past year with the 
observation of $t{\bar t}$ pair production in proton-antiproton collisions at 
the Fermilab Tevatron\cite{ref:cdfdo}.  A deserving question 
is the quantitative reliability of theoretical computations of the total cross 
section, as a function of top mass, based on the main 
production mechanisms in perturbative quantum chromodynamics (pQCD).  In this
paper, we discuss the motivation for incorporating the effects of initial-state 
gluon radiative corrections, and we present our all orders resummation of these
contributions.\cite{ref:Meed}

At lowest order in perturbation theory, two QCD partonic subprocesses 
contribute to $p+{\bar p}\rightarrow t+{\bar t} + X$.  They are 
quark-antiquark annihilation: $q+{\bar q}\rightarrow t+{\bar t}$
and gluon-gluon fusion: $g+g\rightarrow t+{\bar t}$.  Short-distance 
partonic cross sections based on these lowest order
${\cal O}(\alpha_s^2)$ subprocesses and on the next-to-leading  
${\cal O}(\alpha_s^3)$ subprocesses have been investigated 
thoroughly.\cite{ref:ocube,ref:meng}  A full ${\cal O}(\alpha_s^n)$ calculation, 
for $n \ge 4$, does not exist.  The physical top quark cross section is 
obtained 
from a convolution of the perturbative short-distance subprocess cross sections 
with parton distributions that specify the probability densities of the 
quarks, antiquarks, and gluons of the incident $p$ and ${\bar p}$.  Our work
addresses improvements in the reliability of calculations of the subprocess 
cross sections.  

The motivation for this work begins with the observation that the size of 
the ${\cal O}(\alpha_s^3)$ terms in the $q {\bar q}$ and $g g$ partonic cross 
sections are much larger than their ${\cal O}(\alpha_s^2 )$ counterparts in
some kinematic regions, notably in the near-threshold region of small $\eta$. 
The variable $\eta={\hat s}/4m^2-1$, where ${\hat s}$ is the square of the 
energy of parton-parton subprocess and $m$ denotes the mass of the top quark.
Variable $\eta$ measures the ``distance" above the 
partonic production threshold. 

\begin{figure}
\vspace*{13pt}
{\hskip 1cm}\hbox{\epsfxsize6.0cm\epsffile{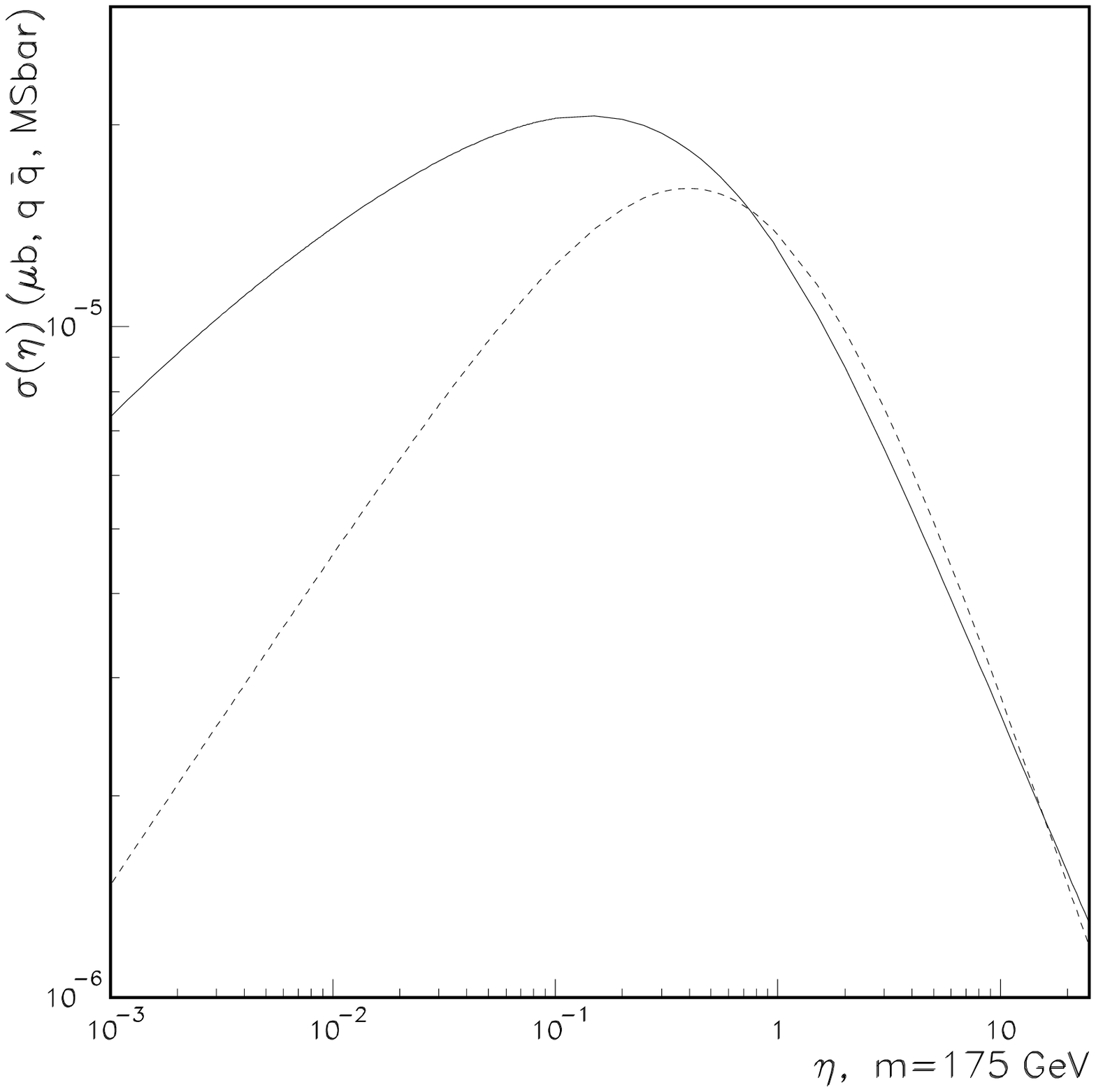}{\hskip 1.6cm}
\epsfxsize6.0cm\epsffile{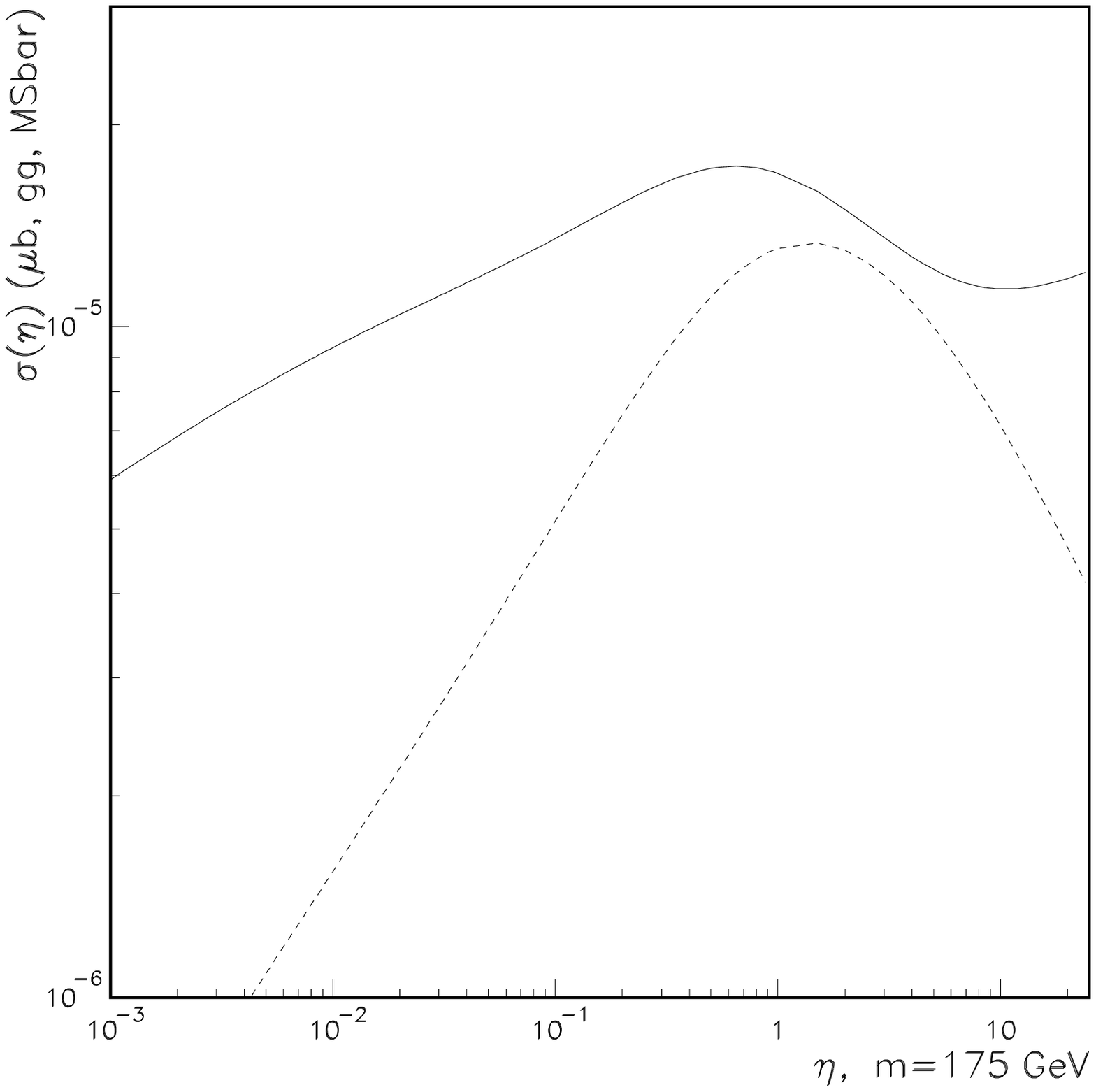}}
\fcaption{The parton-parton cross sections as a function of $\eta$
in the $\overline{MS}$ scheme at $m = 175$ GeV for the subprocesses (a)
$q\bar{q} \rightarrow t\bar{t} X$ and (b)
$gg \rightarrow t\bar{t} X$.  Plotted are the lowest order
Born cross section (dotted line) and the next-to-leading order cross section
(solid line). The QCD scale $\mu = m$.}
\label{fig:fone}
\end{figure}

As shown in Fig. 1, for a top mass of 175 GeV, in both the $q {\bar q}$ and the 
$g g$ channels the size of the 
${\cal O}(\alpha_s^3)$ term exceeds that of the ${\cal O}(\alpha_s^2 )$ term 
for $\eta \simeq$ 0.1, and the ratio grows as $\eta$ decreases.\cite{ref:Meed} 
Therefore, the important 
notion underlying perturbation theory, that successive terms in the 
perturbation series should be smaller, is not valid at small $\eta$, i.e., in 
the region near production threshold.  This region of phase space is important 
for top quark production at the Tevatron.  Owing to the large mass of the top 
quark, relative to the $p{\bar p}$ center of mass energy $\sqrt{s}$, the 
near-threshold region contributes significantly when the convolution integral, 
mentioned above, is done over the full of range of $\eta$.  Confidence in the 
results of a perturbative calculation of the overall $t {\bar t}$ cross section 
requires an appropriate understanding of the origin of the large 
next-to-leading order enhancement of the partonic cross sections near 
threshold.  (In the $g g$ channel, the ratio of the 
${\cal O}(\alpha_s^3)$ and ${\cal O}(\alpha_s^2 )$ terms exceeds unity for 
large $\eta$ also. The $g g$ channel and the large $\eta$ region are important 
for bottom quark production at the Tevatron\cite{ref:keithjohn}, but not for 
top quark production.)

We treat only $t{\bar t}$ pair production.  
Mechanisms for single top production\cite{ref:onetop} are not considered here.
At $m = 175$ GeV and $\sqrt{s}=1.8$ TeV, they contribute a cross 
section about 20\%\ that of the pair production mechanisms.

\section{Gluon Radiation and Resummation}
The origin of the large threshold enhancements in the subprocess cross sections
may be traced to initial-state gluon radiation.\cite{ref:ocube}  After the
cancellation of soft singularities 
between the contributions from real gluon emission and virtual gluon exchange
terms, and proper factorization of collinear divergences, there remain terms
at ${\cal O}(\alpha_s^3)$ that are proportional to $\ell n(1 - z)$.  The
variable $z = 1 - 2k_{g}.p_{t}/m^2$ where $p_{t}$ and $k_{g}$ are the 
four-vector 
momenta of the produced top quark and the gluon radiated into the final state
in the $2 \rightarrow 3$ process.  The limit $z \rightarrow 1$ corresponds 
to zero momentum carried by the gluon. 

The partonic cross section may be expressed generally as
\begin{equation}
\hat{\sigma}_{ij}(\eta,m^2)=\int_{z_{min}}^1
dz\biggl[1+{\cal H}_{ij}(z,\alpha)\biggr]\hat{\sigma}_{ij}'(\eta,m^2,z).
\label{one}
\end{equation}
In the near threshold region,
\begin{equation}
{\cal H}_{ij}(z,\alpha) \simeq 2\alpha C_{ij} \ell n^2 (1-z)
+ \alpha^2\biggl[2C^2_{ij} \ell n^4 (1-z) - {4\over 3} C_{ij} b_2
\ell n^3 (1-z)\biggr] .
\label{two}
\end{equation}
We work in the $\overline{\mbox{MS}}$ factorization scheme in which
the $q$, $\bar{q}$ and $g$ densities and the next-to-leading order
partonic cross sections are defined unambiguously.  In 
Eq.~(\ref{one}), the lower limit of integration $z_{min} = 
1-4(1+\eta)+4\sqrt{1+\eta}$, and $ij\in\{q\bar{q},gg\}$ denotes the initial 
parton channel.  We set $\alpha\equiv \alpha_s(m)/\pi$.  Symbol
$\hat{\sigma}_{ij}'(\eta,m^2,z)=d(\hat{\sigma}_{ij}^{(0)}(\eta,m^2,z))/dz$, 
where $\hat{\sigma}_{ij}^{(0)}$ is the lowest order partonic cross section 
expressed
in terms of inelastic kinematic variables\cite{ref:LSvN} to account for the 
emitted radiation. The integration in Eq.~(\ref{one}) is over the phase space 
of the radiated gluons, parametrized through the dimensionless variable
$z$. In Eq.~(\ref{two}), $C_{ij}$ is the color factor for the $ij$ production 
channel.  

Equation~(\ref{two}) approximates the near-threshold behavior of the partonic
cross section.  It manifests the logarithmic behavior $\ell n(1 - z)$ 
mentioned above.  Explicit calculations\cite{ref:ocube} of the 
complete ${\cal O}(\alpha_s^3)$ cross section provide the 
$2\alpha C_{ij} \ell n^2 (1-z)$ term.  The terms proportional 
to $\alpha^2$ are appropriated from 
${\cal O}(\alpha_s^2)$ computations of massive lepton-pair 
production,\cite{ref:drellyan,ref:Mesterman,ref:Mealvero} 
based on the assumption of universality of leading logarithmic contributions.
As in other hard-scattering processes, where large logarithmic contributions 
are present near threshold, the goal of gluon resummation in  $t {\bar t}$ 
production is to sum the series in $\alpha^{n}\ell n^{2n} (1-z)$ to all orders.
Resummation, studied extensively for massive lepton-pair production,
\cite{ref:drellyan,ref:Mesterman,ref:Mealvero} is important both for 
theoretical understanding of the perturbative process and for stability 
of the quantitative predictions.

In resummation procedures, the large logarithmic contributions are 
exponentiated into a function of the QCD running coupling evaluated at a
variable momentum scale that is a measure of the radiated gluon momentum.  
A straightforward method of resummation for $t{\bar t}$ production was
published a few years ago.\cite{ref:LSvN}  In this approach, the partonic 
cross section of Eq.~(\ref{one}) is replaced in the $\overline{\mbox{MS}}$ 
scheme by the resummed expression
\begin{equation}
\hat{\sigma}_{ij}^{Res}(\eta,m^2,\mu_o)=\int_{z_{min}}^{1-(\mu_o/m)^3}
dz{\rm e}^{E_{ij}^{IRC}(z,m^2)}\hat{\sigma}_{ij}'(\eta,m^2,z),
\label{three}
\end{equation}
where
\begin{equation}
E_{ij}^{IRC}(z,m^2) \propto C_{ij}\alpha((1-z)^{2/3}m^2)\ell n^2 (1-z).
\label{four}
\end{equation}
It is easy to verify that the form of Eqs.~(\ref{one}) and (\ref{two})
is reproduced if ${\rm e}^{E_{ij}^{IRC}}$ is expanded in a power series in 
$\alpha(m^2)$.  A limitation of this method
is that an infrared singularity is encountered in the soft-gluon limit 
$z \rightarrow 1$:  owing to the logarithmic behavior of $\alpha(q^2)$, 
$\alpha(q^2) \propto \ell n^{-1} (q^2/\Lambda_{QCD}^2)$, 
$\alpha((1-z)^{2/3}m^2) \rightarrow \infty$ as $z \rightarrow 1$. 
This divergence of the integrand at the upper limit of integration necessitates
introduction of the undetermined infrared cutoff cutoff $\mu_o$ in 
Eq.~(\ref{three}), $\Lambda_{QCD} \leq \mu_o \leq m$, that serves to prevent 
the integration over $z$ from reaching the Landau pole of the QCD running 
coupling constant.  The 
cutoff has a related effect of eliminating a portion of the integration over 
the partonic subenergy when the convolution with parton densities is
done to obtain the physical cross section.  In terms of $\eta$, the 
convolution is restricted to $\eta \geq \eta_o = {(\mu_o/m)^3}/2$.  The
presence of an extra scale spoils the renormalization group properties of
the overall expression.  Moreover,   
dependence of the resummed cross section on this undetermined cutoff is 
important numerically.\cite{ref:LSvN}  

Laenen {\it et al } furnish their
final predictions in the DIS factorization scheme in which Eqs.~(\ref{three}) 
and (\ref{four}) are modified slightly.  They obtain  
\begin{equation}
\sigma_{t\bar{t}}(m=175\ {\rm GeV})= 4.95^{+0.7}_{-0.4}\ pb\ ,
\label{five}
\end{equation}
based on the assumed values $\mu_o = 0.1m $ for the $q{\bar q}$ channel and 
$\mu_o = 0.25m $ for the $g g$ channel.  Owing to sensitivity to $\mu_o$,
it is difficult to assess the significance of the estimated uncertainties.  

\section{Principal Value Resummation}
The principal-value method of resummation (PVR)\cite{ref:Mesterman}
has an important technical advantage in that it does not depend on arbitrary 
infrared cutoffs, as all Landau-pole singularities are by-passed by a Cauchy 
principal-value prescription.  Because extra undetermined scales are absent, 
the method also permits an evaluation of the perturbative regime of 
applicability of the method, i.e., the region of the gluon radiation phase 
space where perturbation theory should be valid.  The method has been tested 
successfully in massive lepton-pair production.\cite{ref:Mealvero}

To illustrate how infrared cutoffs are avoided in the PVR method,
it is useful to express in moment space the exponent that resums 
the $\ell n (1-z)$ terms:
\begin{equation}
E(n, m^2)= -\int\limits^1_0 dx {{x^{n-1}-1}\over{1-x}}
\int\limits^1_{(1-x)^2} {{d\lambda}\over{\lambda}} g
\left[ \alpha\left( \lambda m^2\right) \right].
\label{six}
\end{equation}
The function $g(\alpha)$ is calculable perturbatively, but, again, the
behavior of $\alpha(\lambda m^2)$ leads to divergence of the integral when 
$\lambda m^2 \rightarrow \Lambda_{QCD}^2$.  To tame the divergence, a cutoff
can be introduced in the integral over $x$ or directly in momentum space, in 
the fashion of Laenen {\it et al }.\cite{ref:LSvN}  In the principal-value 
redefinition of resummation, the singularity is
avoided by replacement of the integral over the real axis $x$ in 
Eq.~(\ref{six}) by an integral along a contour $P$ in the complex plane:
\begin{equation} 
E^{PV}(n, m^2)\equiv -\int\limits_P  
d\zeta {{\zeta^{n-1}-1}\over{1-\zeta}}
\int\limits^1_{(1-\zeta)^2}
{{d\lambda}\over{\lambda}}g\left[\alpha\left(\lambda m^2\right)\right].
\label{seven}
\end{equation}
The function $E^{PV}(n, m^2)$ is finite since the Landau pole singularity 
is by-passed.  In Eq.~(\ref{seven}), all large soft-gluon threshold 
contributions are included through the two-loop running of $\alpha$.  

Equations ~(\ref{six}) and ~(\ref{seven}) have identical perturbative content,
but, when expanded in power series in $\alpha(m^2)$ and in $\Lambda_{QCD}/m$, 
they manifest differences in their inverse power (high-twist) terms.  Since 
the inverse power content is not a 
prediction of perturbative QCD, neither expression is {\it a priori} 
preferable, except for the attractive finiteness of Eq.~(\ref{seven}).  
In our application of principal-value resummation to top quark production, we
choose to use the result only in the region of phase space in which the 
perturbative 
content dominates.  Thus, the high-twist content of Eq.~(\ref{seven}), and
the difference between the high-twist components of Eqs.~(\ref{seven}) and
~(\ref{six}), are not matters of phenomenological significance.  

After inversion of the Mellin transform, the resummed partonic cross sections 
according to PVR, including all large threshold corrections, can be written
in the form of Eq.~(\ref{one}), but with Eq.~(\ref{two}) replaced by 
\begin{equation} 
{\cal H}^{PV}_{ij}(z,\alpha)=\int_0^{\ln({1\over 1-z})}
dx{\rm e}^{E_{ij}(x,\alpha)}
\sum_{j=0}^\infty Q_j(x,\alpha)
\ .\label{afour}
\end{equation}
The leading large threshold corrections are contained in the exponent 
$E_{ij}(x,\alpha)$, a calculable polynomial in $x$.  The functions 
$\{Q_j(x,\alpha)\}$ arise from
the analytical inversion of the Mellin transform from moment space
to the physically relevant momentum space expressed in Eq.~(\ref{afour}). 
These functions are produced by the resummation and are expressed in 
terms of successive derivatives of E:
$P_k(x,\alpha)\equiv\partial^k E(x,\alpha)/k! \partial^k x$.

The functional form of $E_{ij}$ for $t\bar{t}$ production is identical to that 
for $l\bar{l}$ production, except for the identification of the two separate 
channels, denoted by the subscript $ij$.  However, only the {\it leading} 
threshold corrections are universal. Final-state gluon radiation as well as
initial-state/final-state interference effects produce sub-leading logarithmic 
contributions that differ for processes with different final states.  Among
all $\{Q_j\}$ in Eq.~(\ref{afour}), only the very leading one is universal.
This is the linear term in $P_1$, which turns out to be $P_1$ itself.  Since we
intend to resum only the universal leading logarithms, we retain only $P_1$.
Hence, Eq.~(\ref{afour}) can be integrated explicitly, 
and the resummed version of Eq.~(\ref{one}) is
\begin{equation}
\hat{\sigma}_{ij}^{PV}(\eta,m^2)=\int_{z_{min}}^{z_{max}}dz
{\rm e}^{E_{ij}(\ln({1\over 1-z}),\alpha)}
\hat{\sigma}_{ij}'(\eta,m^2,z).\label{threep}
\end{equation}
\begin{figure}
\vspace*{13pt}
{\hskip 1cm}\hbox{\epsfxsize6.0cm\epsffile{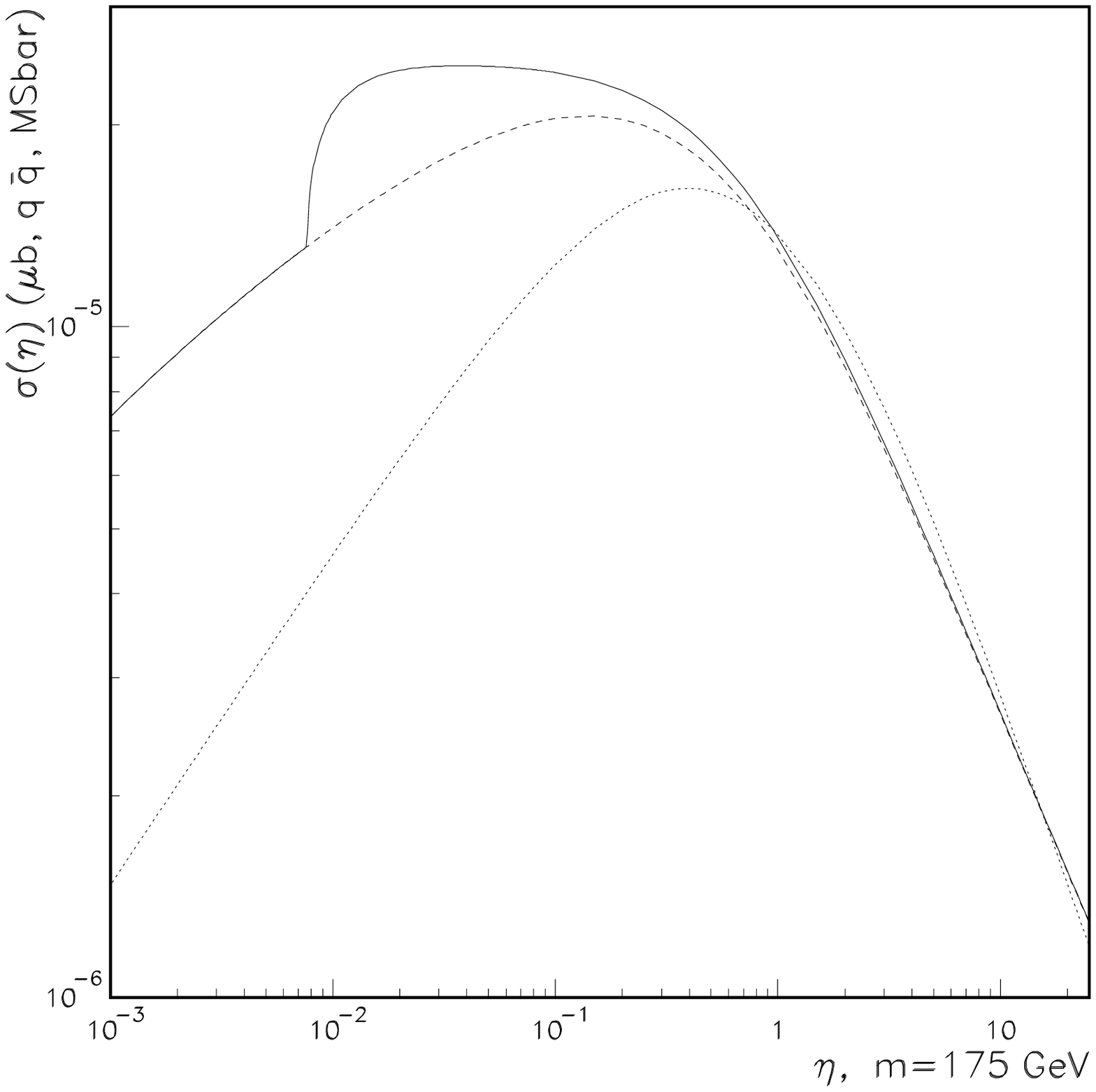}{\hskip 1.6cm}
\epsfxsize6.0cm\epsffile{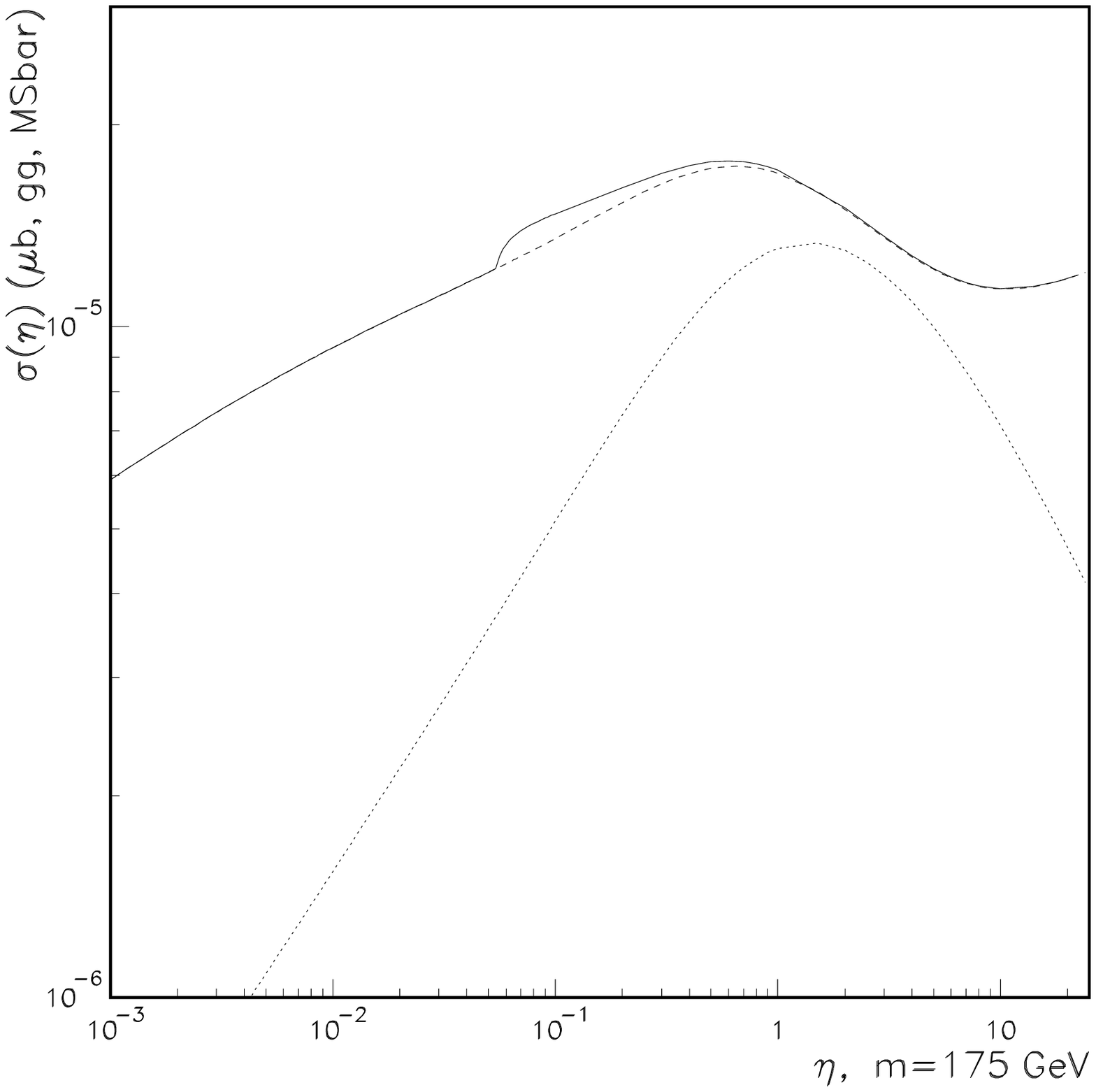}}
\fcaption{The parton-parton cross sections as a function of $\eta$
in the $\overline{MS}$ scheme at $m = 175$ GeV for the subprocesses (a)
$q\bar{q} \rightarrow t\bar{t} X$ and (b)
$gg \rightarrow t\bar{t} X$.  Plotted are the lowest order
Born cross section (dotted line), the next-to-leading order cross section
(dashed line), and the resummed cross section (solid line). The QCD scale
$\mu = m$.}
\label{fig:ftwo}
\end{figure}
The upper limit of integration in Eq.~(\ref{threep}) is set by the boundary
between the perturbative and high-twist regimes.  To characterize a 
region in moment space as high-twist, one must convert to momentum 
space through inversion of the Mellin 
transform, Eq.~(\ref{afour}). Specification of the boundary is realized
by the constraint that all $\{Q_j\},\ j\ge 1$ be small compared to
$Q_0$.  This constraint can be shown to correspond to 
\begin{equation}
P_1\left(\ln\left({1\over 1-z}\right),\alpha\right)<1\ .
\label{pertmom}
\end{equation} 
As remarked above, we accept only the perturbative content of
principal-value resummation, and our cross section is evaluated accordingly.  
Specifically, we use Eq.~(\ref{threep}) with the upper limit of integration, 
$z_{max}$, calculated from Eq.~(\ref{pertmom}).  The upshot is an effective 
threshold cutoff on the integral over the scaled subenergy variable
$\eta$, reminiscent of that introduced by Laenen {\it et al }, but one that
is calculable, {\it not} arbitrary.  In our case, the cutoff restricts the 
region of applicability of resummation to the part of phase space in which the 
perturbative content of Eq.~(\ref{afour}) is the dominant content.  For the
top mass $m$ = 175 GeV, we determine that the perturbative regime is
restricted to $\eta \geq$ 0.007 for the $q{\bar q}$ channel and 
$\eta \geq$ 0.05 for the $gg$ channel.  The difference reflects the larger
color factor in the $gg$ case.  (One could attempt to apply Eq.~(\ref{threep}) 
all the way to $z_{max} = 1$, i.e., to $\eta =$ 0, 
beyond the perturbative regime of Eq.~(\ref{pertmom}), but
one would then be using a {\it model} for non-perturbative
effects, the one suggested by PVR, far beyond the knowledge justified
by perturbation theory.) 

Our final result does not rely critically on the PVR method to 
by-pass infrared renormalons and associated problems, precisely because 
we restrict application to the perturbative regime. In this regard, the 
presence of arbitrary infrared cutoffs in other resummation methods is 
superfluous, as all necessary information about infrared sensitivity
(i.e., the perturbative regime) can be obtained by examining the 
perturbative asymptotic properties of the resummation functions.  

The resummation procedure includes only the leading threshold $\ell n^2(1 - z)$
piece of the full ${\cal O}(\alpha_s^3)$ calculation.  To restore the full
content of the complete next-to-leading order calculation, 
$\hat{\sigma}_{ij}^{(0+1)}$, we define our final resummed 
cross sections for each production channel through the improved prediction
\begin{equation}
\hat{\sigma}^{\rm final}_{ij}(\eta,m^2)=
\hat{\sigma}^{PV_{\rm pert}}_{ij}(\eta,m^2)+
\hat{\sigma}^{(0+1)}_{ij}(\eta,m^2)-
\hat{\sigma}^{(0+1)}_{ij}(\eta,m^2)\bigg|_{PV}\ .
\label{afive}
\end{equation}
The last term in Eq.~(\ref{afive}) is the part of the next-to-leading order 
partonic cross section included in the resummation. 
In Fig.2 we present the resummed partonic cross sections in 
the $q\bar{q}$ and $gg$ channels at $m=175$ GeV. We also
show the lowest order and next-to-leading order
counterparts.  The three curves differ substantially in the partonic threshold
region $\eta<1$, with the final resummed curve exceeding
the other two.  Below $\eta\simeq 0.007$ in the $q\bar{q}$ channel
and $\eta\simeq 0.05$ in the $gg$ channel, our resummed 
cross sections become identical to the next-to-leading order
cross sections, a consequence of our decision to restrict 
the resummation to the perturbative domain.  Above $\eta\simeq 1$, our 
resummed cross sections are
essentially identical to the next-to-leading order cross sections,
as is to be expected since the near-threshold enhancements that concern us
in this paper are not relevant at large $\eta$. 

\section{Calculations at $\sqrt{s}=1.8$ TeV and $\sqrt{s}=2$ TeV}
In the remainder of this report, we  present our results for the physical
inclusive total cross section for $t\bar{t}$ production for a top-mass
range $m\in\{150,250\}$ GeV, including a discussion of the remaining theoretical
uncertainties.  We obtain the physical cross section by convoluting 
Eq.~(\ref{afive}) with CTEQ3M parton densities\cite{ref:cteq} 
and adding the results from both channels.  In Fig. 3 we show the top-mass 
dependence of the physical cross section for 
$p\bar{p}\rightarrow (t\bar{t})X$ at $\sqrt{s}=1.8$ TeV.

\begin{figure}
\vspace*{13pt}
{\hskip 1.6cm}\epsfxsize12.0cm\epsfbox[0 140 570 650]{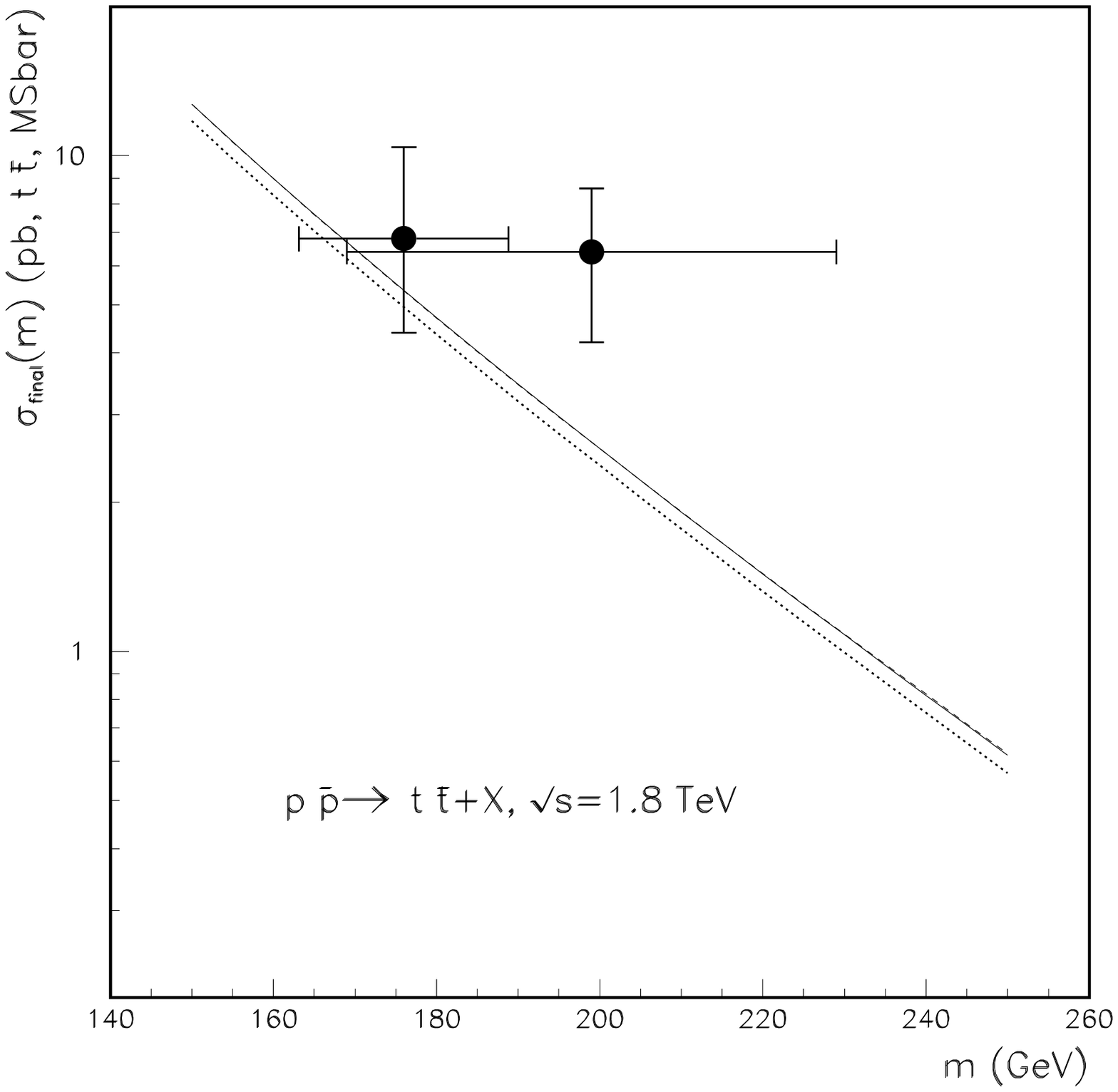}
\fcaption{Physical cross section for $p\bar{p}\rightarrow (t\bar{t})X$
at $\sqrt{s}=1.8$ TeV as a function of top mass. Data from the
CDF and D0 collaborations\cite{ref:cdfdo} are plotted. Shown are
calculations for three choices of the scale $\mu/m=0.5$ (dashed),
$1$ (solid), and $2$ (dotted).}
\label{fig:fthree}
\end{figure}

As illustrated in Fig. 4, the behavior of the physical cross section as a 
function of the renormalization/factorization scale $\mu$ is mild in the 
range $\mu/m\in\{0.5,2\}$. (The next-to-leading order result in Fig. 4 
differs somewhat from that shown in our published paper.\cite{ref:Meed}  Owing 
to a compiler error, the next-to-leading order curve in the published figure is 
incorrect.)  We consider the variation of the 
cross section in the range $\mu/m\in\{0.5,2\}$ to be a reasonable measure of 
the theoretical perturbative uncertainty.  Over this range, the band
of variation of the strong coupling strength $\alpha_s$ is a 
generous $\pm10$\%\ .  We determine
\begin{equation}
\sigma^{t\bar{t}}(m=175\ {\rm GeV},\sqrt{s}=1.8\ {\rm TeV})=
5.52^{+0.07}_{-0.45}\ pb\ .
\label{predone}
\end{equation}
\begin{figure}
\vspace*{13pt}
{\hskip 1.5cm}\epsfxsize12.0cm\epsfbox[0 370 550 700]{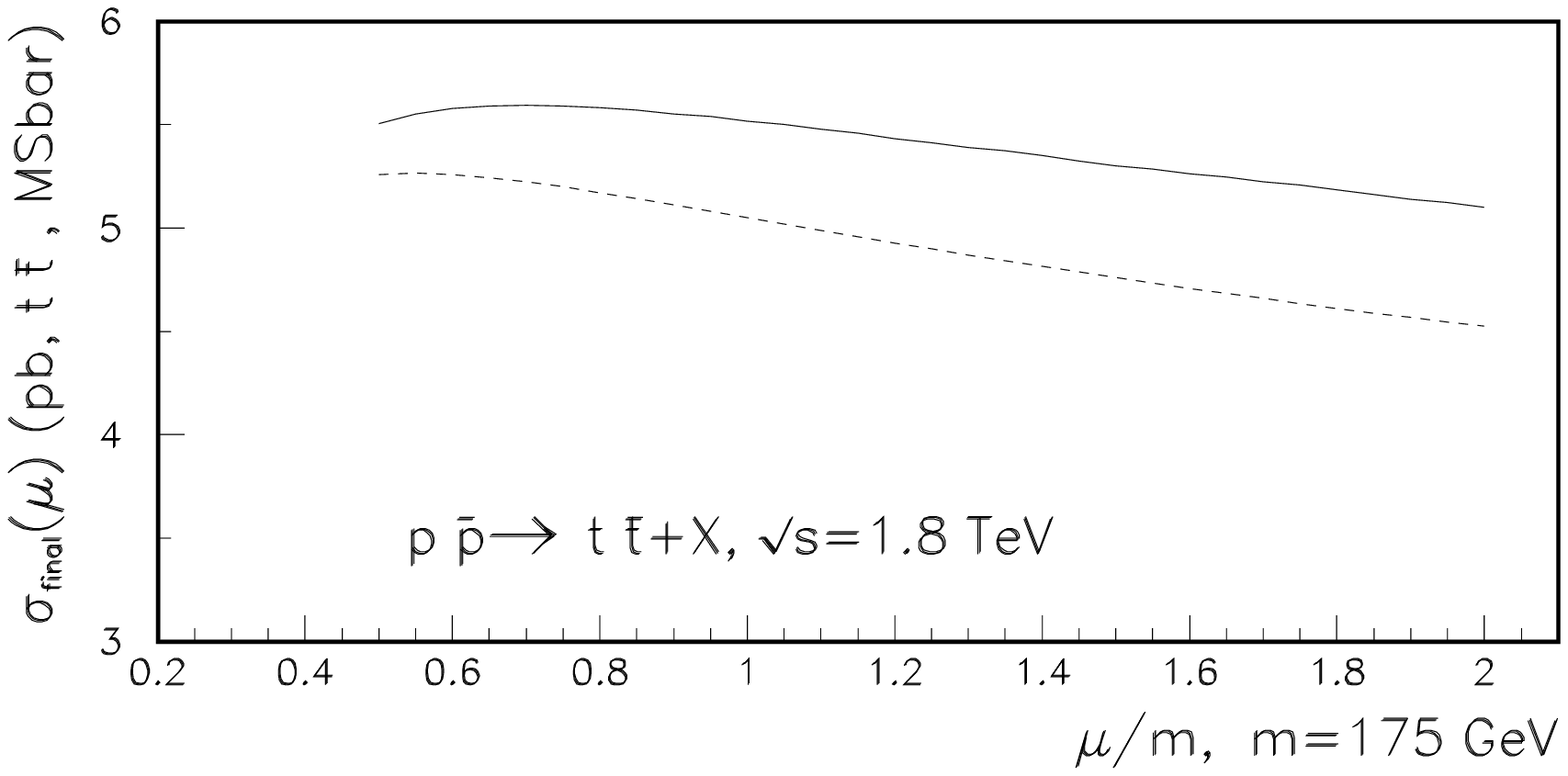}
\fcaption{Plot showing the calculated dependence of the final resummed
cross section on $(\mu/m)$ for $t\bar{t}$ production
at $m = 175$ GeV and  $\sqrt{s} = 1.8$ TeV.
Shown also is the next-to-leading order result (dashed curve).}
\label{fig:ffour}
\end{figure}
We define the central value (5.52 $pb$) to be that obtained with $\mu/m=1$.
The upper and lower limits correspond to the maximum
and minimum values of the cross section in the range $\mu/m\in\{0.5, 2\}$.
The cross section is insensitive to the choice of parton densities.
Repeating the same analysis with the MRS($A^\prime$) densities\cite{ref:mrsa},
we obtain
\begin{equation}
\sigma^{t\bar{t}}(m=175\ {\rm GeV},\sqrt{s}=1.8\ {\rm TeV})=
5.32^{+0.08}_{-0.41}\ pb\ .
\label{predtwo}
\end{equation}
The central values in Eqs.~(\ref{predone}) and ~(\ref{predtwo}) are about 
10\%\ larger than that of Laenen {\it et al }, Eq.~(\ref{five}), but within the 
quoted uncertainties.  Our calculated cross sections fall within the current 
uncertainty bands of the CDF and D0 experiments.\cite{ref:cdfdo}

The bands of perturbative uncertainty quoted in Eqs.~(\ref{predone}) 
and ~(\ref{predtwo}) are relatively narrow. On the other hand, 
we noted in discussing Eq.~(\ref{threep}) that $z_{\rm max}<1$, meaning 
that there is a reasonable range of $\eta$ near threshold in 
which perturbative resummation does not apply. Perturbation theory
is not justified in this region. Correspondingly, further strong interaction
enhancements of the $t\bar{t}$ cross section may arise from physics
in this region. We know of no reliable way to estimate the size of 
such non-perturbative effects and, therefore, cannot include such 
uncertainties in the estimates of the {\it perturbative} uncertainty
of Eqs.~(\ref{predone}) and (\ref{predtwo}).
\begin{figure}[htb]
{\hskip 1.6cm}\epsfxsize12.0cm\epsfbox[0 140 570 650]{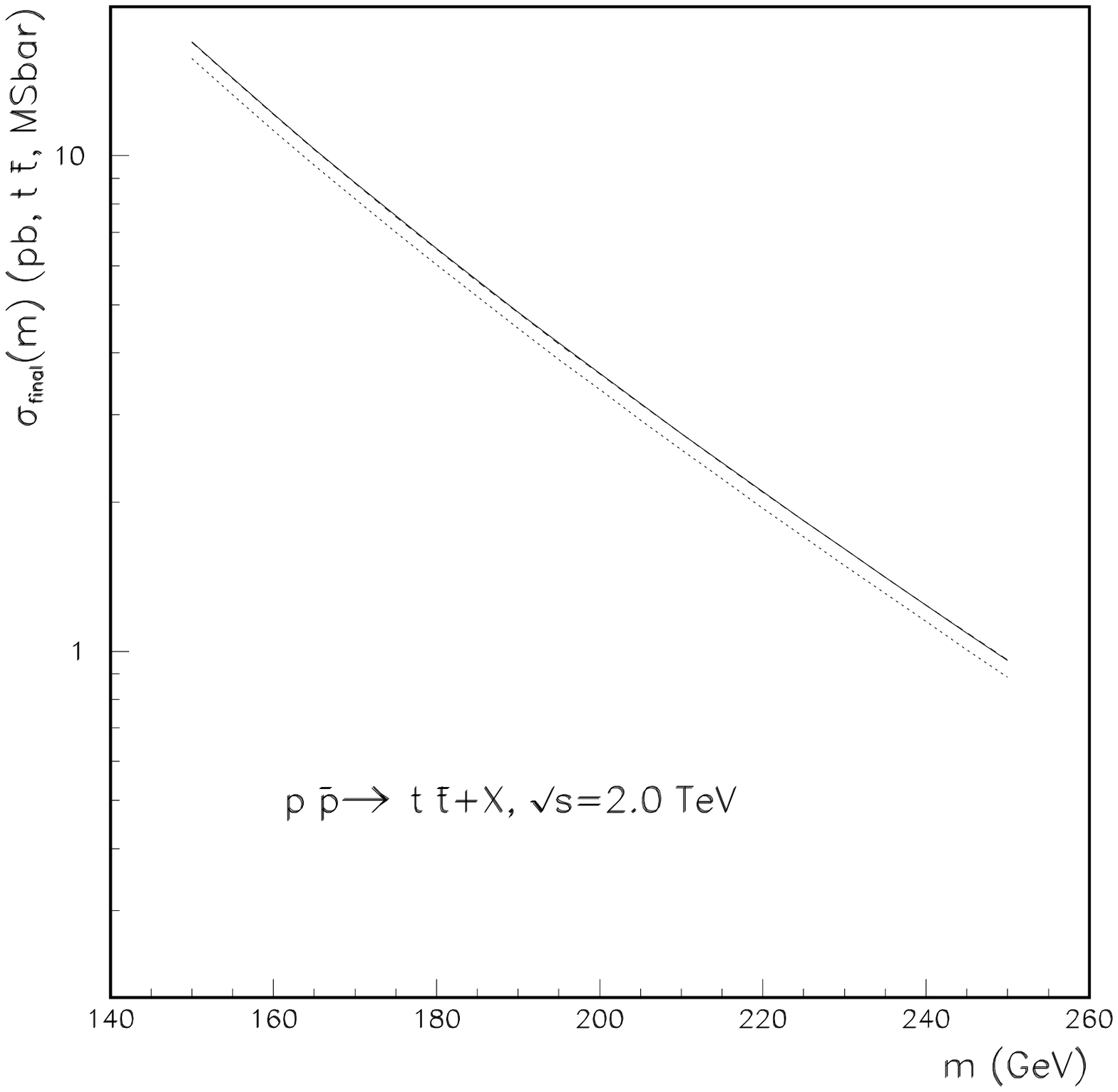}
\fcaption{Physical cross section for $p\bar{p}\rightarrow (t\bar{t})X$
at $\sqrt{s}=2$ TeV as a function of top mass.  Shown are
calculations for three choices of the scale $\mu/m=0.5$ (dashed),
$1$ (solid), and $2$ (dotted).}
\label{fig:ffive}
\end{figure}

In Fig. 5 we show the top-mass dependence of
the physical cross section for $p\bar{p}\rightarrow (t\bar{t})X$ at
the slightly larger energy $\sqrt{s} = 2$ TeV.  We predict
\begin{equation}
\sigma^{t\bar{t}}(m=175\ {\rm GeV},\sqrt{s}=2\ {\rm TeV})=
7.56^{+0.10}_{-0.55}\ pb\ .
\label{predthree}
\end{equation}
At $m = 175$ GeV, the value of the cross section at $\sqrt{s} = 2$ TeV is 
about 37\%\ greater than that at $\sqrt{s}= 1.8$ TeV.

\section{LHC Predictions}
Turning to $pp$ scattering at the energies of the Large Hadron Collider (LHC)
at CERN, we note a few significant differences from $p\bar{p}$ scattering at the
energy of the Fermilab Tevatron.  The dominance of the $q {\bar q}$ production
channel at the Tevatron is replaced by $g g$ dominance at the LHC.  Owing to
the much larger value of $\sqrt{s}$, the near-threshold region in the subenergy
variable is relatively less important, reducing the significance of 
initial-state soft
gluon radiation.  Lastly, physics in the region of large $\sqrt{\hat{s}}$, 
where straightforward next-to-leading order QCD is also inadequate, may become
significant for $t\bar{t}$ production at LHC energies.  Using the approach 
described in 
this paper, focussed on the resummation of initial-state gluon radiation,
we present predictions in Fig. 6 for LHC energies of 10 and 14 TeV. We estimate
\begin{equation}
\sigma^{t\bar{t}}(m=175\ {\rm GeV},\sqrt{s}=14\ {\rm TeV})=760\ pb\ .
\label{predfour}
\end{equation}
\begin{figure}[htb]
\vspace*{13pt}
{\hskip 1.6cm}\epsfxsize12.0cm\epsfbox[0 140 570 650]{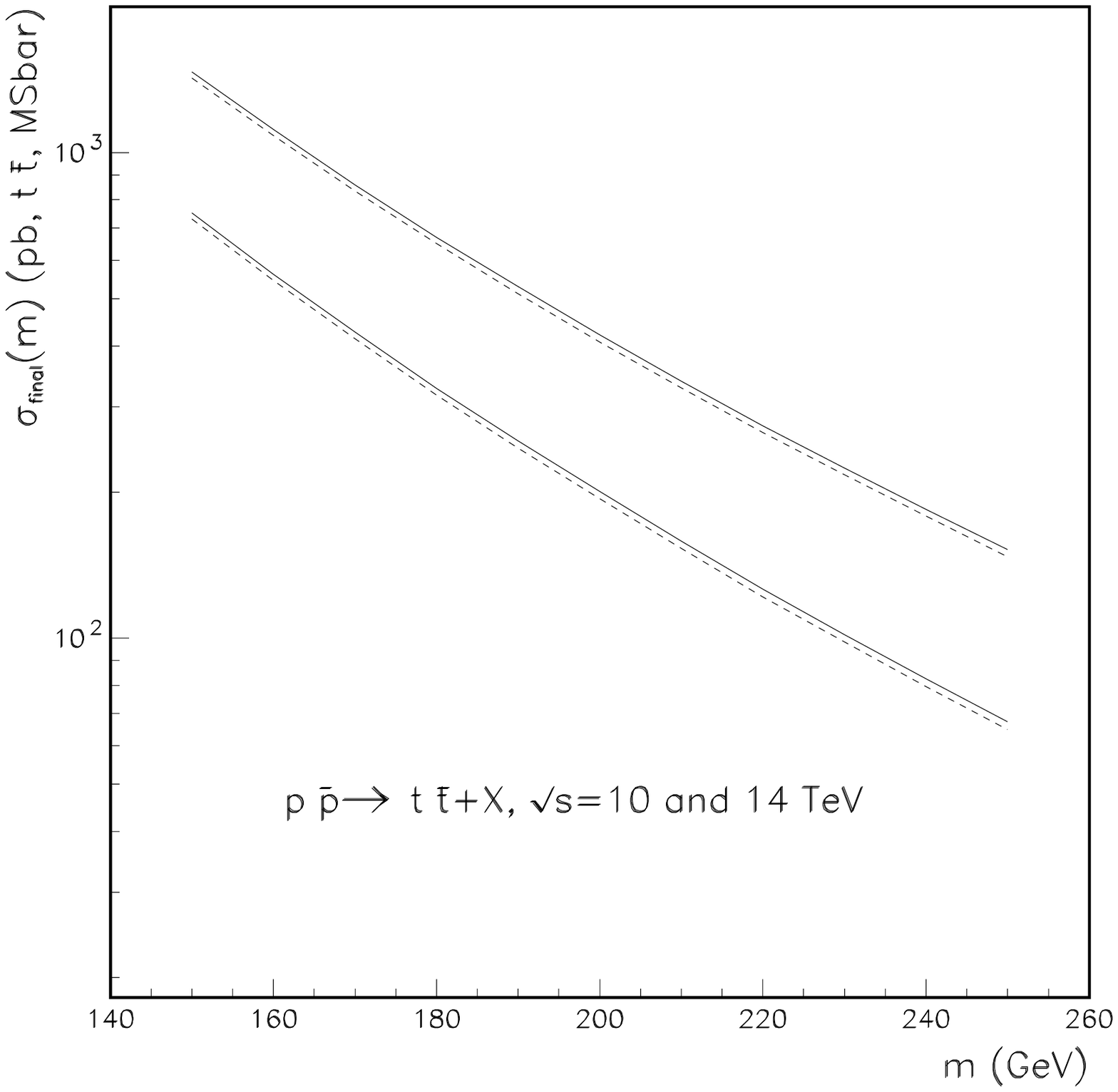}
\fcaption{Cross section for $p p\rightarrow (t\bar{t})X$
at $\sqrt{s}= $10 and 14 TeV as a function of top mass.  Shown are
the next-to-leading order (dashed) and resummed (solid) calculations for 
the scale choice $\mu/m= 1$.}
\label{fig:fsix}
\end{figure}

\section{Summary and Discussion}
In summary, we present a calculation of the total cross section for 
$t\bar{t}$ pair production in perturbative QCD including resummation of
initial-sate gluon radiation to all orders in $\alpha_s$.  Two advantages
of the principal-value method of resummation are the well-defined
perturbative domain of applicability and the absence of arbitrary infrared
cutoffs.  Both $q {\bar q}$ and $g g$ production channels are included in the
calculation.  At $\sqrt{s} =$ 1.8 TeV, our final resummed cross section is 
approximately 10\%\ greater than the pure next-to-leading order result and in
agreement with data. We provide predictions for the cross section as a function
of top mass at $\sqrt{s} =$ 2, 10, and 14 TeV.  

We remark that the resummation method developed here and applied to $t\bar{t}$ 
pair production is relevant in other situations in which dynamics probes
the near-threshold region in the scaled subenergy variable.  An example of
current interest at $\sqrt{s} =$ 1.8 TeV is the production of hadronic jets 
that carry large transverse momentum.\cite{ref:cdfpt}

\section{Acknowledgements}
ELB is pleased to acknowledge valuable comments from Stanley Brodsky and John
Ellis, and he is grateful to Chao-Hsi Chang and Tau Huang for their warm 
hospitality in Beijing.  This work was supported by the U.S. Department of 
Energy, Division of High Energy Physics, contract W-31-109-Eng-38.

\medskip

\end{document}